\documentclass[amsfonts,amssymb,amsmath,prl,preprint]{revtex4}

\usepackage{times}
\usepackage{graphicx}
\usepackage{bm}
\usepackage{color}

\definecolor{pink}{rgb}{1,1,0} 
\definecolor{red}{rgb}{1,0,0}
\definecolor{yellow}{rgb}{1,1,0}
\definecolor{orange}{rgb}{1,0.5,0}
\definecolor{white}{rgb}{1,1,1}

\newcommand{\be}{\begin{equation}}
\newcommand{\ee}{\end{equation}}

\begin{document}

\title{Impurity crystal in a Bose-Einstein condensate }

\author{ David C. Roberts$^1$ and Sergio Rica$^2$}  

\affiliation{$^1$ Theoretical division and Center for Nonlinear Studies, LANL, Los Alamos, NM, USA  \\ $^2$ Laboratoire de Physique Statistique, CNRS-Ecole normale sup\'erieure, 24 rue Lhomond, 75005 Paris, France. 
} 
\begin{abstract} 
We investigate the behavior of impurity fields immersed in a larger
condensate field in 1, 2, and 3 dimensions.  We discuss the
localization of a single impurity field within a condensate and note
the effects of surface energy.  We derive the functional form of the
attractive interaction between two impurities due to mediation from
the condensate.  Generalizing the analysis to $N$ impurity
fields, we show that within various parameter regimes a crystal of
impurity fields can form spontaneously in the condensate.
Finally, the system of condensate and crystallized impurity
structure is shown to have nonclassical rotational inertia, which is
characteristic of superfluidity, {\it i.e.} the system can be seen to exhibit supersolid
behavior.
\end{abstract}

\date{\today}

\maketitle

{\it Introduction.--} Recent observations in solid helium \cite{ss} have generated enormous
speculation about the existence of supersolidity, first predicted almost forty years ago \cite{al}.  This issue remains controversial, partly due to
the difficulty of investigating the helium system both theoretically
and experimentally.  Trapped atomic Bose-Einstein condensates (BECs),
which exhibit superfluid behavior, are a promising alternative system
to probe supersolidity as one can not only tune the atomic
interactions, but also calculate many of their properties from first
principles.  In this paper we present a model of small impurity fields
immersed within a larger condensate (impurities in Bose liquids have been considered in many contexts \cite{bli}) that mimics some of the
fundamental properties of supersolids, namely that it breaks the
translational symmetry of the Hamiltonian as well as exhibits
off-diagonal long range order as evidenced by nonclassical
rotational inertia (NCRI).  Unlike previous supersolid proposals in
dilute BECs \cite{ricapomeau},  here the crystal
scale emerges spontaneously from the system rather than being
externally imposed.

The present model is realizable with current experimental technology.
For instance, the distinct impurity fields can be produced by
utilizing isotopes or by transferring atoms from a larger condensate
via Raman pulses into different atomic levels in an atomic trap.  One
can also explore parameter space by tuning the coupling constants via
a magnetic and optical Feshbach resonance. 

{\it The model.--} In this paper, we consider a large BEC denoted by the field $\psi$ coupled to $N$ small distinguishable impurity fields denoted by $\chi_k$.  To permit an uncluttered description of the system's nontrivial properties, we shall assume that all impurities interact with the same coupling constants. (We briefly  consider more realistic systems in the discussion.) Our system is therefore governed by the following Hamiltonian: 
\begin{eqnarray}
\label{energy}
H = \int \left[ \frac{1}{2} | \nabla \psi |^2 + \frac{1}{2}  |\psi|^4 + \lambda |\psi |^2   \sum_{k=1}^{N} | \chi_k |^2   +  \sum_{k=1}^{N}  \frac{1}{2m} | \nabla \chi_k |^2 + \frac{\gamma_0}{2}  \sum_{k=1}^{N} | \chi_k |^4 + \gamma  \sum_{j<k}^{N} | \chi_j |^2| \chi_k |^2  \right]  d {\bm x}
\label{ham}
\end{eqnarray}
where $\lambda$ is the coupling of the condensate to the impurity fields, $\gamma_0$ is the self-interaction of the impurity fields, and $\gamma$ is the interaction between impurity fields.  If $\gamma_0=0$ this Hamiltonian describes distinguishable impurities that do not self-interact. We shall restrict ourselves to positive parameters. In dilute BECs these coupling constants are directly proportional to the atomic scattering length with a proportionality constant of $4 \pi \hbar^2/m$.   As mentioned above, these scattering lengths are in principle tunable via magnetic fields in current experiments.

From eq.(\ref{ham}), the system's dynamics are governed by $N+1$ coupled nonlinear Schr\"odinger equations:
\begin{eqnarray}
i \partial_t \psi \, &=& \, -\frac{1}{2} \Delta \psi \, + \,  |\psi|^2 \psi +\lambda \psi  \sum_{k=1}^{N} | \chi_k |^2  \label{nls}\\
i \partial_t \chi_k\, &=& \, -\frac{1}{2m} \Delta \chi_k \, + \, \gamma_0  |\chi_k|^2 \chi_k + \gamma \chi_k  \sum_{j\neq k}^{N} | \chi_j |^2 +  \lambda | \psi|^2  \chi_k \label{nls.imp}
\end{eqnarray}
where $\Delta$ is the Laplacian in $D$ spatial dimensions.  In this system there is particle conservation of the large condensate field $N = \int  |\psi|^2  d^D{\bm x} $ and of each impurity field $n_k = \int  |\chi_k|^2  d^D{\bm x} $, and we assume $n_k \ll N$.  The total energy (\ref{energy}) and the total linear momentum $ {\bm P } = {\rm Im} \int  \psi^*{\bm \nabla} \psi \, d^D {\bm x} +  {\rm Im} \sum_{k=1}^N \frac{1}{m}  \int  \chi_k^*{\bm \nabla} \chi_k \, d^D {\bm x} $ are also conserved.  

{\it Instabilities of the uniform state and collapse of the system.--} Let us consider the nontrivial structures that impurity fields can generate in a condensate.  To determine the critical point at which these structures emerge, we begin with the uniform miscible state, which is stable for certain values of the coupling constants.  Ignoring surface tension effects (discussed below), one can show that the uniform state is stable if and only if the $(N+1)\times(N+1)$ matrix, $\mathcal M$,  defined by the quadratic form  of the potential energy in (\ref{ham}), is positive semidefinite, i.e. if all eigenvalues of $\mathcal M$ are positive \cite{robertsueda}.  One can also prove that if $\mathcal M$ is negative semidefinite the system will experience finite-time collapse \cite{robertsnewell}.  Furthermore, because the density is non-negative, one can extend the criteria in \cite{robertsnewell} and show that the system would also experience a finite-time collapse if  $\mathcal M$ is conegative or positive subdefinite.  We will focus mainly on the regime where the uniform state has a modulational instability (implying that at least one eigenvalue of $\mathcal M$ is negative) and, since the coupling constants are assumed positive, we do not consider situations proven to exhibit finite-time collapse.

The $N+1$ eigenvalues of $\mathcal M$ are $\gamma_0-\gamma$, which is $N-1$ degenerate, and $\frac{1}{2}(1 +(N-1) \gamma + \gamma_0)  \pm  \frac{1}{2} \sqrt{ (N-1) \gamma + \gamma_0 -1  ) ^2 + 4 N \lambda^2}$.  Therefore, a modulational instability occurs if $\gamma>\gamma_0$ or if $\lambda > \frac{{\sqrt{(N-1)\,\gamma  +{\gamma_0}}}}{{\sqrt{N}}}$, highlighting the difference between the system's two distinct types of modulational instabilities.  In the first regime, $\gamma>\gamma_0$, each impurity localizes individually whether or not the system is phase separated from the condensate.  In the second regime, the 
condensate phase separates from the impurity fields.  

{\it A single impurity: Interpolating between self-localization and phase separation.--} We now discuss the system of a small impurity field embedded in and interacting with a large condensate.  A variational argument will show that there is a critical value of the coupling parameter (dependent on surface tension effects) between the condensate and impurity beyond which the impurity self-localizes. 

The ground-state wave functions in 1, 2, and 3 dimensions should possess the following characteristics: that $\chi (r)$ be localized such that $\chi (r) \rightarrow 0 $ as $r  \rightarrow \infty$, and its mass be fixed such that $n_k = C_D  \int  |\chi_k|^2  r^{D-1}  d r   ,$ where  $C_D = \frac{2\,{\pi }^{\frac{D}{2}}}  {{\Gamma}(\frac{D}{2})} $ is the surface of a unit sphere in $D$ spatial dimensions; that there be a depletion where the impurity is positioned, and $ \psi (r)  \rightarrow \psi_0 = cte$ as $r  \rightarrow \infty$.    The energy to be minimized is the Hamiltonian (\ref{ham}) for one impurity relative to the energy of uniform state assuming $\psi (r) =  \psi_0$. 

Consider the real, normalized trial functions of the form $\chi(r) = \sqrt{n_k} \sqrt{\frac{\alpha^D} {C_D \, {\mathcal N}_D} } f({\alpha }\, r)$ and $\psi (r) = \psi_0\left( 1 - a \,  \chi (r) ^2  \right)$, where $a$ and $\alpha > 0$ are variational parameters determined by minimizing the energy estimates (note that any real finite system does not allow $\alpha \rightarrow 0$) and $f(r)$ is localized such that $f(r) \rightarrow 0$ as $r \rightarrow \infty$.  The normalization constant is ${\mathcal N}_D = \int_0^\infty  f(x)^2 x^{D-1}\,dx $.  With a formal expansion for the energy, one can easily show that $a\approx \frac{\lambda}{2 \psi_0^2}  $ in the limit  $\alpha\rightarrow 0$ .

The energy  in $D$ dimensions is thus {\it bounded} by
\begin{equation} \label{Ealpha}
E\leq E(\alpha) =
\epsilon_0  \, {\alpha }^2+ 
  \epsilon_1\, {\alpha }^D
   + \epsilon_2  \,{\alpha }^{2 + D}   + 
 \epsilon_3  \,{\alpha }^{3\,D}
 \end{equation}
 where the constants $\epsilon_0,\, \epsilon_2$, and $\epsilon_3$ are all positive numbers (detailed calculations will be presented elsewhere), but $ \epsilon_1 = k_1  \left({\gamma_0} - \lambda^2 \right) n_k^2$ (where $k_1$ is positive)may change its sign if $\lambda^2 > \gamma_0$. 

  It is important to emphasize that the variational approach only gives us an upper bound on the ground-state energy.   If for a nonzero $\alpha$ the lowest variational energy is negative (recall the energy for the nearly uniform state is positive and approaches zero as $\alpha \rightarrow 0$), then we know the uniform state is unstable and can be reasonably  sure that self-localization has occurred.  However, if the lowest variational energy is non-negative, we cannot determine whether or not self-localization occurs.  Nonetheless, this expression of energy as a function of the variational parameter $\alpha$, eq. (\ref{Ealpha}), does give useful insight into the localized impurity solution in 1-, 2-, or 3-dimensional space.  When $\epsilon_1>0$, the energy is a monotonically increasing function of $\alpha$, implying that the nearly uniform ground state ($\alpha \rightarrow 0$) minimizes the energy. 
 
For $D=1$, the dominant term at small $\alpha$ is $ \epsilon_1\, {\alpha }$.  Thus when $\epsilon_1$ is negative, i.e. $\lambda^2>\gamma_0$, a supercritical transition occurs from a homogeneous state to a localized impurity state. 
 
For $D=2$,  the first and second terms are of the same order. As is the case for $D=1$, there is a second-order transition towards a localized impurity state if $\lambda^2 > \gamma_0 + \frac{k_1}{m n_k} $, with $k_1 = 4 C_2\, {\mathcal N}_2  \frac{\int_0^\infty  f'(x)^2 \, x \,dx  }{\int_0^\infty  f(x)^4 \, x \,dx }$. The instability of the homogeneous state is shifted from the bulk condition described above.  This shift has a simple interpretation: the presence of a $1/m$ factor means that this term comes from the kinetic energy of the impurity $\int \frac{1}{2 m}  |\nabla \chi_k|^2 $, so the shift is created by the curvature $\alpha$ of the localized structure --it is a kind of surface tension. Furthermore, $\frac{\int_0^\infty  f'(x)^2 \, x \,dx  }{\int_0^\infty  f(x)^4 \, x \,dx }$, as the ratio of an interface energy to a bulk energy, bears the hallmark of a surface-tension effect.

For $D=3$,  the situation is more subtle.   As discussed above, the variational argument states that a localized solution exists if the minimum energy in (\ref{Ealpha}) is negative for some critical value, $\alpha_c$.  A necessary (but not sufficient) condition for a negative-energy  ground state  is $\lambda^2 > \gamma_0$.  We can make a more precise estimate as follows:
because the energy turns negative for some range of $\alpha >0$ and the energy expansion grows quadratically near $\alpha\approx 0$ (since $\epsilon_0>0$ in (\ref{Ealpha})), it is sufficient (though not necessary) that $E(\alpha_c)$ is a minimum and that $E(\alpha_c)\le 0$.  The critical line defining onset of self-localization is given by $E(\alpha_c)=0$ and $E'(\alpha_c)=0$; on one side we can say that a localized structure exists, on the other side we cannot be certain.
This line maybe written in a parametric way by the following:
$
 6 \beta ^7+2 \beta ^3={\varepsilon_0 \varepsilon_3^{3/4}}/{\varepsilon_2^{7/4}}\quad \&\quad 7 \beta ^6+3 \beta ^2+{\varepsilon_1\sqrt{\varepsilon_3}}/{\varepsilon_2^{3/2}}=0,
$
where $\beta \equiv \alpha_c \left({\varepsilon_3}/{\varepsilon_2}    \right)^{1/6}$ is the parametrization. 
In the gaussian approximation one gets the upper bound ${\rm sup}\left\{  \left( \frac{9 \pi}{2}\right)^{3/4}    \frac{1}{\sqrt{ m n \psi_0} }  ,  \frac{7^{7/10} \pi^{3/5}}{2^{17/20}}   \frac{1}{m^{3/5} (n \psi_0) ^{2/5}} \right\}$ in the case of $\gamma_0=0$,  which is about 35\% higher than the numerical result \cite{eddy}.

{\it Condensate-mediated attraction.--}  Here we discuss the interaction between $N$ dilute localized impurity fields and derive the effect of the perturbed background condensate on the interaction.  We will show that the interaction has an  attractive tail mediated by the condensate in addition to the hard-core repulsion arising from the repulsion between impurities.   

Let us consider $N$ self-localized impurity fields (having satisfied the conditions given in the previous section) that weakly modify the uniform condensate, i.e. $\psi = \psi_0 + \psi_1({\bm r})$.  This assumption is valid if the impurity is either only weakly localized or sufficiently distant.  The latter situation is of particular relevance here as we are interested in deriving the long-range attractive tail of the interaction. In this approximation, the condensate wave function is linear and may be solved with the aid of a Green's function in $D$-spatial dimensions.  The solution is well known and can be written explicitly as
\begin{equation}G^{(D)}({\bm x}-{\bm x}')  = \left\{ 
		\begin{array}{ll}
 \frac{1}{4 \psi_0} e^{-2\psi_0 | {\bm x}-{\bm x}'|} & \quad {\rm} in \, D=1 \\
  K_0(2\psi_0 | {\bm x}-{\bm x}'|) & \quad {\rm} in \, D=2 \\
    \frac{e^{-2\psi_0 | {\bm x}-{\bm x}'|}}{ | {\bm x}-{\bm x}'|} & \quad {\rm} in \, D=3\\
				\end{array}
\right.
\label{yukawa}
\end{equation}
where $K_0$ is the modified Bessel's function.  We can thus determine $\psi_1({\bm x})$.  Assuming the condensate to be only weakly modified, eliminating the terms that do not depend explicitly on the impurity fields, and explicitly omitting the self-interaction energy of the impurities which is given by \cite{eddy}, the interacting energy of the system of $N$ impurities becomes 
\begin{eqnarray}
E =  \frac{1}{2}
 \sum_{i \neq k}^{N}   \int U({\bm x}-{\bm x}')    | \chi_i ({\bm x}')|^2  | \chi_k ({\bm x})|^2 \,d^{D}{\bm x}\,  d^{D}{\bm x} '     
\label{interatingenergy2}
\end{eqnarray}
where \begin{equation}
U({\bm x}_i-{\bm x}_k)= \gamma \delta^{(D)}({\bm x}_i-{\bm x}_k) -   \frac{4 \lambda^2 \psi_0^2 }{C_D}   G^{(D)}({\bm x}_i-{\bm x}_k).  
\label{interactingenergy3}
\end{equation}

Let us consider localized impurities ($\alpha|x_k -x_i|  \gg 1$). One may approximate the $k$-th impurity field by $ | \chi_k ({\bm x})|^2= n_k \delta^{(D)}({\bm x}-{\bm x}_k)$ where ${\bm x}_k$ is the position of the $k$-th impurity.  In  this case $E=  \frac{1}{2}\sum_{i\neq k}^{N} n_i n_k  \, {\mathcal U}(|{\bm x}_i-{\bm x}_k|) $, where the
interaction potential between the two impurities is given by ${\mathcal U}({\bm x} )= U({\bm x} )$ of (\ref{interactingenergy3}). (The yukawa attractive tale for similar systems in 3D were discussed in \cite{fermiyukawa}) 
The first $\delta$-interacting energy term in (\ref{interactingenergy3}) is only a crude estimation. The full expression (\ref{interactingenergy3}) says that a bound state exists but the equilibrium distance is zero. In reality the equilibrium distance is on the order of the size of the impurity. Introducing the trial function $  \chi_k ({\bm x})=  \sqrt{n_k} \sqrt{\frac{\alpha^D} {C_D \, {\mathcal N}_D} } f(\alpha|{\bm x}-{\bm x}_k|)$ with $f(s)=e^{-s^2}$, into the interaction energy (\ref{interatingenergy2}), for $D=3$ on finds ${\mathcal U}(|{\bm x}|)= \gamma \frac{\alpha^3}{\pi^{3/2}} e^{-\alpha^2|{\bm x}|^2} -   \frac{ \lambda^2 \psi_0^2 }{\pi}     \frac{e^{-2\psi_0 |{\bm x}| }}{ |{\bm x}|}$.  The second term was arrived at by assuming $\alpha \gg 2 \psi_0$, i.e. that the impurity is more localized than the Yukawa interaction range $1/(2\psi_0)$. The equilibrium distances come from this energy.

\begin{figure}[htc]
\begin{center}
\centerline{ \includegraphics[width=7cm]{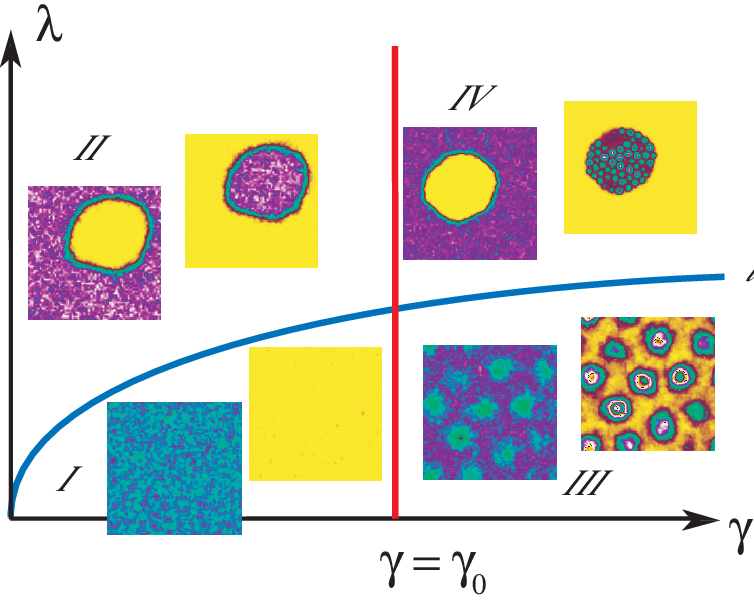}   }
\caption{ \label{fig1} 
 Phase diagram.  The vertical line represents  $\gamma=\gamma_0$ while the curve {\it i} represents $\frac{{\sqrt{(N-1)\,\gamma  +{\gamma_0}}}}{{\sqrt{N}}}$. All numerical simulations of Eqns. (\ref{nls},\ref{nls.imp}) are in two space dimensions in a $64 \times 64 $ periodic plane.  The initial conditions were uniform miscible states plus  small fluctuations. In each sector the left image plots the condensate density $|\psi|^2$ while the right, plots $\sum_{k=1}^N |\chi_k|^2$. In {\it III} the initial masses are $\int |\psi|^2 d^2{\bm x} = 4096$ and  $\int |\chi_k|^2 d^2{\bm x} = 40.96 $, $\lambda=0.2$, $\gamma=1$ and $\gamma_0=0$ while in   {\it IV} the initial masses are $\int |\psi|^2 d^2{\bm x} = 4096$ and  $\int |\chi_k|^2 d^2{\bm x} = 40.96 $, $\lambda=2$, $\gamma=1.5$ and $\gamma_0=0$.    Note that, as discussed in the text, the surface tension effects will shift the boundaries.
 }
\end{center}
\end{figure}

{\it N impurities | crystallization.--} We now discuss the system in which we have $N$ interacting impurity fields within a much larger condensate.  We showed in the previous section that within a certain parameter regime the localized impurities exhibit a tunable hard-core attractive weak interaction in any $D$ with like particles, a system that has been studied in the context of many diverse physical systems (for 3D classical particles see e.g. \cite{yukawa}).   We will show that these impurities crystallize in two regimes of condensate-impurity interaction.  Figure \ref{fig1} is a phase diagram showing four distinct regimes.  In phase I, the condensate and impurities are miscible.  In phase  {\it II}, the condensate phase separates from the impurities, which remain together and miscible with each other in a bubble.  No crystallization occurs in either phase.  Crystallization {\it does} occur in phases {\it III} and {\it IV}.  In phase  {\it III}, the impurities remain immersed in the condensate; in phase {\it IV} they phase separate, forming a crystal within a bubble within the condensate.

\begin{figure}[htc]
\begin{center}
\centerline{{\it a)} \includegraphics[width=3.7cm]{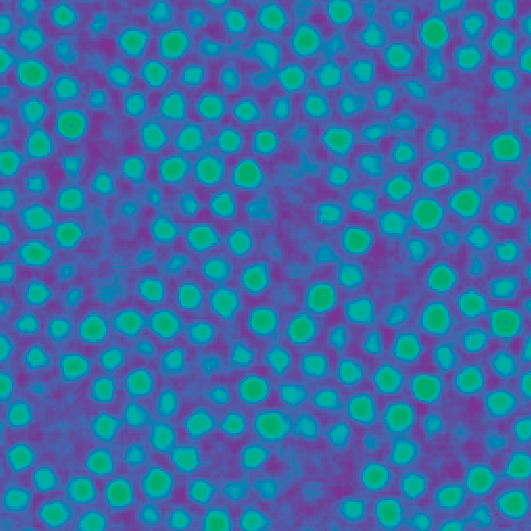} \quad {\it b)}  \includegraphics[width=3.7cm]{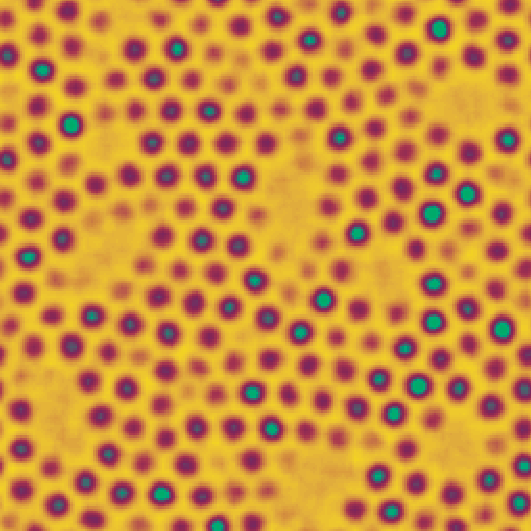}}
\caption{ \label{figyukawa}
Numerical simulation of Eqns. (\ref{nls},\ref{nls.imp}) for $N=36$ impurity fields in a $128 \times 128 $ periodic plane. Here the initial masses are $\int |\psi|^2 d^2{\bm x} = 16384 $ and  $\int |\chi_k|^2 d^2{\bm x} = 200 $, $\lambda=0.4$, $\gamma=1$ and $\gamma_0=0$. Left {\it a)}, plots the condensate density $|\psi|^2$ while the right, {\it b)} plots $\sum_{k=1}^N |\chi_k|^2 $.  Inhomogeneous regions, i.e. where miscible states coexist with the
crystalline state, indicate that the system has not yet reached an
equilibrium state.
 }
\end{center}
\end{figure}

In phase {\it III}, the impurities are attracted through the interaction energy (\ref{interatingenergy2}).  However, there is a repulsive hard core, and an equilibrium distance is expected.  The system crystallizes as is clearly shown through numerics (see Fig. \ref{figyukawa}).

In phase  {\it IV}, there is phase separation of the condensate from the impurities such that the impurities are confined in a kind of bubble within which the condensate density is near zero.  At the same time, inside the bubble there is also phase separation of the impurity fields from each other.  The interaction between the impurities is no longer has an attractive tail because $\psi_0\approx 0$ in (\ref{yukawa}).  The condensate essentially plays the role of container since any impurity that begins to stray from the ensemble becomes more strongly attracted to it as more condensate seeps in  between, causing an attractive interaction between the stray impurity and the ensemble.  Finally, crystallization arises because of the ``pressure'' of the condensate boundary and the short-distance repulsion, see Fig. \ref{fig1}-{\it IV}. 

In some sense, the system in phase  {\it IV} is at a special limit when the condensate is not coupled to the impurity fields.  Therefore, the case $\lambda =0$ deserves special attention. The condensate can be described by the pure nonlinear Schr\"odinger equation, which is quite well understood. The impurity fields, however, evolve.  Impurity fields that overlap store potential energy and will tend to repel each other; all the more so as $\gamma$ increases.  If the impurity fields $|\chi_k|^2$ are localized about ${\bm R}_k$ with a size $\delta$ then, in the limit $\gamma\gg \gamma_0$, the potential energy is dominated by $ \frac{\gamma}{2} \sum_{i\neq k}  \int |\chi_i({\bm x}) |^2 |\chi_k({\bm x})|^2  d^D{\bm x}$.  Minimizing the overlap will minimize the potential energy.  In some sense, the energy may be approximated by the superposition of two-body repulsive  interactions. In a finite box there is an upper bound on the separation between the impurities.  One may invoke a close-packing argument to find the crystal structure that minimizes the energy of the impurity ensemble, sustained only by the external pressure from the condensate at the boundaries. 

The periodic case where all $N$ impurity fields have the same mass $n_{*}$ provides an interesting, solvable example. One can use the minimization approach to determine $\delta$ as a function of the large parameter $\gamma$. To begin, let us assume that the impurity field vanishes exactly outside the ball of radius $\delta$, that there is no overlap, and that only the nearest neighbor(s) affect the interaction energy.  The minimum of the total energy $H = \sum_{k=1}^{N}  \int  \frac{1}{2m} | \nabla \chi_k |^2    d {\bm x}$ can be determined from the solutions of the Hemholtz equation $-\frac{1}{2m} \Delta \chi_k = \varepsilon \chi_k$ for a ball $V_\delta({\bm R}_k)$ with a Dirichlet boundary condition on $\partial V_\delta({\bm R}_k) $ \cite{amandine}.  One can perturb this to determine the effect of a slight overlap.  The variational parameter $\delta$ will determine the optimal configuration that balances the kinetic energy of a pulse and the interaction energy due to a small overlap of impurity fields. The one-dimensional case in a periodic domain is a solvable example that we shall present elsewhere.

{\it Nonclassical rotational inertia.--} We have shown that in a certain parameter regime impurities immersed in a condensate crystallize.  We will now show that the crystallized phases-{\it III \& IV}  still behaves as a superfluid ({\it i.e.} it has NCRI), and thus exhibits supersolid characteristics. 

Following \cite{leggett}, consider a cylindrical system of volume  $V$, rotating uniformly about its primary axis of rotation $\hat {\bm e}_1$. For a small angular rotation the system will possess an energy ${\mathcal E} = {\mathcal E}_0 +  \Delta {\mathcal E}$ where ${\mathcal E} _0 $ is the ground state energy and $\Delta {\mathcal E} =\frac{1}{2} {I}^{eff}\omega^2$, $ {I}^{eff} $ being the effective and measurable moment of inertia tensor around the $\hat {\bm e}_1$ axis. Deviations of this tensor from the rigid body rotation tensor $I^{RB}$ are called the nonclassical rotational inertia fraction.

Under rotation, the phases of the condensate and impurity fields are no longer uniform and the increase of energy is $\Delta {\mathcal E}  =   \frac{1}{2}  \int \left(  \rho_0({\bm x} )\, | \nabla \phi |^2  +  \sum_{k=1}^{N}  \frac{1}{m}  \rho_k({\bm x} ) | \nabla \phi_k |^2   \right)  d {\bm x}$
where $\psi({\bm x}) =\sqrt{ \rho_0({\bm x} )} e^{i \phi} $, $ \chi_k({\bm x}) = \sqrt{\rho_k({\bm x} )} e^{i\phi_k}$ are the nonuniform density of the respective ground states, and the phases $\left\{\phi , \phi_k\right\}$ satisfy the no matter flux boundary conditions:
$\hat n \cdot {\bm \nabla}\left\{\phi , \phi_k\right\} =\omega  \hat n \cdot (\hat e_1 \times {\bm r})$. That $\Delta {\mathcal E}$ is a quadratic form in $\omega$ is evident since $\phi $ and $\phi_k$ must be proportional to $\omega$ due to this boundary condition. The explicit form of ${I}^{eff}$ and the pre-factor needs deeper consideration (see \cite{ss1} for details).
The minimization of $\Delta {\mathcal E}$ leads to continuity equations and boundary conditions that may be solved using the method called homogenization, which splits cleanly the large (system size) and small (impurity size) scales and provides effective average quantities.  
 However, it is not possible to obtain a closed expression of $I^{eff} $ in terms of the local density $ \rho_0({\bm x} )$ and $ \rho_k({\bm x} )$.  Nevetherless, it is possible to show that $I^{eff} \leq I^{RB}$ and that $ I^{eff}$ is proportional to the superfluid density $ \varrho^{ss}$. In one dimension, the problem is exactly solved | a direct calculation leads to Leggett's formula \cite{leggett} :  $\varrho^{ss}=  \left(\frac{1}{V} \int_{V}  \frac{d {x}}{\rho_0({x} ) }  \right)^{-1} + m\sum_{k=1}^N \left(\frac{1}{V} \int_{V}  \frac{d {x}}{\rho_k({x} ) }  \right)^{-1} .$ 

Because the impurity fields are localized, their wave functions decay fast in space so the contribution to the superfluid density from the impurity fields becomes negligible.  Therefore, to leading order the supersolid properties are due to the large condensate; while modulation by the impurities causes the NCRI fraction to be less than unity, the impurities themselves do not contibute to the NCRI.

{\it Discussion.--} Although we have considered systems with identical impurity interaction constants, it is interesting to consider a more realistic system of many distinguishable impurities (perhaps different atomic levels in a single atom) with different coupling constants.  Assuming the system is stable and possesses a modulation instability (with the constraints on coupling constants discussed above), we expect this system to break the translational symmetry of the Hamiltonian as systems discussed in this paper and to possess a NCRI.  However, rather than forming a regular periodic crystal, we expect it to form a giant coherent molecule or amorphous solid with properties directly related to the interaction constants.

\end{document}